# Towards 20 T Hybrid Accelerator Dipole Magnets

P. Ferracin, G. Ambrosio, D. Arbelaez, L. Brouwer, E. Barzi, L. Cooley, L. Garcia Fajardo, R. Gupta, M. Juchno, V. Kashikhin, V. Marinozzi, I. Novitski, E. Rochepault, J. Stern, A. Zlobin, N. Zucchi

*Abstract*—The most effective way to achieve very high collision energies in a circular particle accelerator is to maximize the field strength of the main bending dipoles. In dipole magnets using Nb-Ti superconductor the practical field limit is considered to be 8-9 T. When $Nb_3Sn$ superconductor material is utilized, a field level of 15-16 T can be achieved. To further push the magnetic field beyond the $Nb_3Sn$ limits, High Temperature Superconductors (HTS) need to be considered in the magnet design. The most promising HTS materials for particle accelerator magnets are Bi2212 and REBCO. However, their outstanding performance comes with a significantly higher cost. Therefore, an economically viable option towards 20 T dipole magnets could consist in an "hybrid" solution, where both HTS and $Nb_3Sn$ materials are used. We discuss in this paper preliminary conceptual designs of various 20 T hybrid magnet concepts. After the definition of the overall design criteria, the coil dimensions and parameters are investigated with finite element models based on simple sector coils. Preliminary 2D cross-section computation results are then presented and three main layouts compared: cos-theta, block, and common-coil. Both traditional designs and more advanced stress-management options are considered.

## I. Introduction

**H**IGH-ENERGY proton-proton circular colliders are among the most powerful tool for direct discovery of new particles and interactions [1]. In a circular accelerator, the most effective way to achieve very high collision energies is to maximize the field strength of the main bending dipoles. So far, particle accelerators like the Tevatron [2], HERA [3], RHIC [4], and LHC [5] have used Nb-Ti, a low temperature superconducting (LTS) material, in their main dipole magnets to achieve fields up to the 9 T level. In the High-Luminosity LHC, for the first time, superconducting magnets based on $Nb_3Sn$, also an LTS material, and operating at a 11-12 T field level will be installed in the LHC interaction regions to increase the collision rate [6]. In parallel, R&D programs in Europe, as part of the FCC collaboration [7], and in the US, as part of the national Magnet Development Program (MDP) [8], are developing superconducting magnets aiming at bore fields of 15 to 16 T, currently considered as the practical limit for $Nb_3Sn$ accelerator magnets [9].

To further push the magnetic field of the dipole magnets beyond the $Nb_3Sn$ limits, High Temp. Superconductors (HTS) need to be considered in the magnet design. For accelerator magnets, the most promising HTS materials currently under consideration are Bi2212 [10] and REBCO [11]. However, their outstanding performance still comes with a significantly higher cost than $Nb_3Sn$. Therefore, an economically viable option of 20 T dipole magnets has to involve a "hybrid" approach, where HTS materials are used in the high field part of the coil with so-called "insert coils", and $Nb_3Sn$ and Nb-Ti superconductors are adopted in the lower field region with so-called "outsert coils". Preliminary design studies of 20 T hybrid dipole magnets were carried out in 2005 [12] and in 2014-2016 [13]-[15], whereas a full HTS option was analyzed in 2018 [16]. In 2015, a 20 T hybrid block-type design was presented by G. Sabbi, *et al.* in [17]. A hybrid magnet was recently attempted by inserting a REBCO coil inside the FRESCA2 dipole magnet [18]. Finally, REBCO inserts based on Roebel cables were fabricated and tested as part of the EUCARD2 Collaboration [19]-[20].

In this paper a preliminary conceptual magnetic design analysis of a 20 T hybrid dipole magnet for particle accelerators implementing $Nb_3Sn$ and HTS coils (Nb-Ti is not considered in this study) is presented. First, a description of the superconducting material properties and of the criteria used in the different designs is provided. Then, an analysis of coil size using sector coils, followed by the magnetic analysis of 3 types of coil layouts, 1) cos-theta, with and without stress management, 2) block-type, with and without stress management, and 3) common-coil, are presented. The work focuses on the field level and margins, and only preliminary considerations related to coil mechanics will be provided; a more detailed analysis will follow.

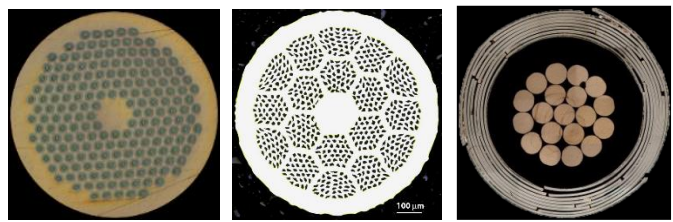

Fig. 1. Cross-sections, not in scale, of $Nb_3Sn$ (left, 0.85 mm $\varnothing$) and Bi2212 (center, 0.7 mm $\varnothing$) composite wires produced by Bruker-OST, and a REBCO CORC wire (right, 3.4 mm $\varnothing$) by ACT LLC.

This work was supported by the U.S. Department of Energy, Office of Science, Office of High Energy Physics, through the US Magnet Development Program (*Corresponding author: Paolo Ferracin*).

P. Ferracin. D. Arbelaez, L. Brouwer, L. Garcia Fajardo, and M. Juchno are with Lawrence Berkeley National Lab, Berkeley, CA 94720, USA (e-mail: pferracin@lbl.gov).

G. Ambrosio, E. Barzi, V.V. Kashikhin, V. Marinozzi, I. Novitski, and A.V. Zlobin are with Fermi National Accelerator Laboratory, Batavia, IL 80510 USA.

R. Gupta is with BNL, Upton, NY 11973-5000, USA.

L. D. Cooley is with the Applied Superconductivity Center, National High Magnetic Laboratory, Tallahassee, FL 32310, USA

E. Rochepault is with IRFU, CEA, Univers Paris-Saclay, Paris F-91191, France.

J. Stern, and N. Zucchi are with TUFTS University, 419 Boston Ave, Medford, MA 02155, USA.

## II. Conductor Parameters

An example of the superconducting materials considered for this analysis is shown in Fig. 1, with a Nb$_3$Sn Rod restack Process (RRP) and a Bi2212 strands produced by Bruker-OST, and a REBCO CORC wire produced by APC LLC. For each material, the engineering current density $j_e$ used in the computations, where $j_e$ is the critical strand current divided by the strand cross-section area, is plotted in Fig. 2. A summary of the superconductor properties is provided in Table I.

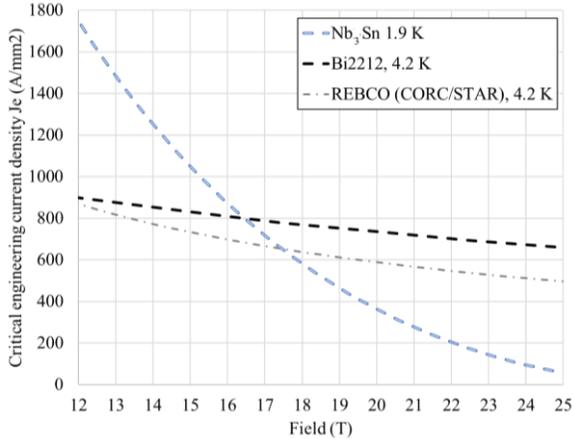

Fig. 2. Critical engineering current density ($j_e$ = I$_{strand}$/A$_{strand}$) assumed for the computations for Nb$_3$Sn, Bi2212 and REBCO CORC/STAR round wires.

### A. Nb$_3$Sn

Nb$_3$Sn strands have been used in both R&D magnets and accelerator magnets in diameters ranging typically from 0.7 to 1.1 mm [9], [21]. The cross-section of a 0.85 mm strand with 132/169 RRP architecture is shown in Fig. 1, left. Nb$_3$Sn Rutherford cables have been fabricated with up to 60 strands, with widths ranging from 7.8 to 26.2 mm and thicknesses within 1.2-2.0 mm. As an example, the cross-section of the Nb$_3$Sn Rutherford cable fabricated for the MQXF project is shown in Fig. 3, top [22]. For the numerical computations we assumed a critical current density in the Nb$_3$Sn (virgin strand) of 3000 A/mm$^2$ at 12 T and 4.2 K. Considering a 1.1 Cu/Non-Cu ratio, this corresponds to a $j_e$ of 870 A/mm$^2$ at 16 T, 1.9 K, including 5% of cabling degradation. Assuming a 0.150 mm thick insulation one obtains a ratio between $j_o$ and $j_e$ of 0.67 (using the MQXF insulated cable parameters [22]), where $j_o$ is the ratio of the cable current to the insulated cable area.

### B. Bi2212

The cross-section of a Bruker-OST Bi-2212 strand, produced with the powder-in-tube method is shown in Fig. 1, center. The Bi2212 filaments are combined in 18 bundles (of 55 filaments each) and embedded in a silver matrix. The reaction process is performed following an overpressure heat treatment (OPHT), which allowed a significant improvement of the $j_e$ in recent years. For the analysis in this paper, we assumed a $j_e$ of 740 A/mm$^2$ at 1.9 K and 20 T, a value obtained in short samples for the racetrack sub-scale coils [23]. As Nb$_3$Sn, the Bi2212 strands will be used in form of Rutherford cable. In terms of strand and cable dimensions, we assumed that the same ranges as for the Nb$_3$Sn can be achieved.

TABLE I
SUPERCONDUCTOR STRAND AND CABLE PROPERTIES

| Parameter | Unit | Nb$_3$Sn | Bi2212 | REBCO |
|---|---|---|---|---|
| Strand diameter | mm | 0.7-1.1 | 0.7-1.1 | 1.2-4.0 |
| Cable width | mm | 7.8-26.2 | 7.8-26.2 | NA |
| Cable thickness | mm | 1.2-2.0 | 1.2-2.0 | NA |
| $j_e$ (at 1.9 K, 16 T) | A/mm$^2$ | 870 | 800 | 700 |
| $j_e$ (at 1.9 K, 20 T) | A/mm$^2$ | 360 | 740 | 590 |
| $j_o / j_e$ | | 0.67 | 0.67 | 0.54 |

### C. REBCO

The CORC (Conductor on Round Core) wire from ACT, shown in Fig. 1, right and described in details in [24], is fabricated by winding several REBCO tapes around a Cu core, with a total wire diameter of about 3-4 mm. These wires were successfully tested in Canted Cos-theta (CCT) coils at LBNL in 2019 [25]. Smaller diameters, of 1.3-2.0 mm, have been achieved with STAR (Symmetric Tape Round) wires described in [26]. In both cases, we assumed for the computations a $j_e$ of 590 A/mm$^2$ at 1.9 K and 20 T. In order to increase the total current carrying capabilities, two paths are being explored: an increase of the number of tapes and of the diameter of the wire, or a combination of smaller wires in a multi-strand cable. For this analysis we considered the option of a large CORC wire with a 6.5 mm diameter, individually powered and inserted in grooves, as shown in Fig. 3, bottom. In this configuration, the $j_o / j_e$ ratio reduces, from 0.67 typical for Rutherford cable to 0.54.

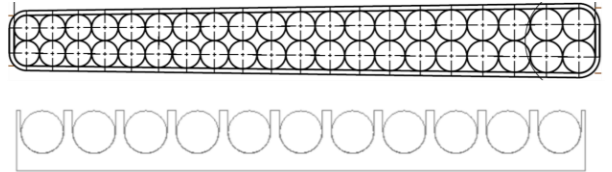

Fig. 3. Figures not in scale. Top: cross-section of the insulated Nb$_3$Sn Rutherford cable for MQXF (18.7 mm wide); bottom: schematic view of CORC/STAR wires in grooves (total width 55 mm).

## III. Design Criteria

The criteria defined in the conceptual design are summarized in Table II. The magnet shall be able to produce 20 T in a 50 mm clear aperture with at least 15% of load-line margin. This means that the "short-sample" bore field, i.e. the bore field achieved when the magnet reaches the current limits established by the conductor critical surface, is 23.5 T. The design shall have all the geometrical harmonics field below 5 units at the nominal field and at 2/3 of the aperture radius (magnetization effects are not included at this stage). All the coils shall be powered in series (a condition which impacts both the magnetic design and the quench protection system), and the hot spot temperature at quench shall be limited to 350 K, both in the Nb$_3$Sn [27] and in the HTS coils [28]-[30]. In terms of stress limits, for the Nb$_3$Sn we chose 150 and 180 MPa at 293 K and 1.9 K respectively, consistently with results published in [31], whereas for the HTS we defined a preliminary and more conservative value of 120 MPa [32], [33]. Both quench protection and coil stress levels will not be considered in the preliminary design studies presented in this paper (Section V).

TABLE II
DESIGN CRITERIA ON MAGNET PARAMETERS

| Parameter | Unit | |
|---|---|---|
| Aperture | mm | 50 |
| Operational temperature | K | 1.9 |
| Operational bore field $B_{0\_op}$ | T | 20 |
| Load-line margin | % | ≥15 |
| Geometrical harmonics (20 T, $R_{ref}$=17 mm) | unit | <5 |
| Maximum Nb$_3$Sn coil eq. stress (293 K) | MPa | 150 |
| Maximum Nb$_3$Sn coil eq. stress (1.9 K) | MPa | 180 |
| Maximum HTS coil eq. stress (293K, 1.9 K) | MPa | 120 |
| Maximum hot spot temperature | K | 350 |

## IV. ANALYSIS WITH SECTOR COILS

For a preliminary investigation of the overall coil size, load-line margins and ratio between LTS and HTS coil area, we performed a magnetic analysis using sector coils. The study follows the same approach as that described in [34], [35], where the superconducting coil is simulated with 60° sector, and with a uniform $j_o$. The cross-sections of the analyzed sector coils analyzed are shown in Fig. 4, where the outserts (in red) are assigned the properties of the Nb$_3$Sn and the inserts (in grey) the properties of Bi2212. The analysis includes an iron yoke (not shown in the figure), which starts at 25 mm from the coil outer radius of the coil and has a thickness of 250 mm. The Bi2212 coil has an aperture of 50 mm, and the two coils are dimensioned so that with a 20 T bore field they both operate at 85% of their limit on the load-line.

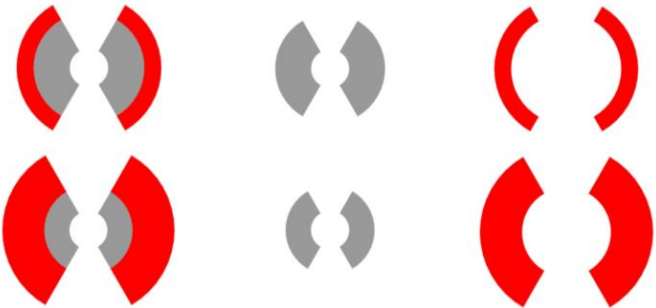

Fig. 4. Sector coils with Bi2212 coil as insert and Nb$_3$Sn coil as outsert. Case I (top): 20 T dipole generated with equal $j_o$ in both insert and outsert (left). Case II (bottom): 20 T dipole generated with $j_o$ in the outsert 33% lower than in the insert (left). For both cases, the stand-alone coils are also shown (center and right).

The main parameters of the different cross-sections are listed in Table III, and the corresponding load-lines are plotted in Fig. 5. Two cases were studied: in the first one (Case I, Fig. 4 top) we applied the same uniform $j_o$ to both insert and outsert coils. This can be seen in Fig. 5, where the two markers in the Case I load-line pointed to a $j_o$ = 570 A/mm$^2$ in both Nb$_3$Sn and Bi2212 for a bore field of 20 T, and to a $j_o$ = 570 A/mm$^2$ for the short sample condition. For this case, the overall coil width is 69 mm, and the majority of it (47 mm) is given by the Bi2212 coil. If the insert and outsert coils are tested in stand-alone with iron (Fig. 4, center and right), at 85% of their load-line limits they generate a bore field of 16.0 T in 50 mm aperture and 10.7 T in 144 mm aperture respectively.

TABLE III
SECTOR COILS PARAMETERS FOR 20 T BORE FIELD

| Parameter | Unit | Case I | Case II |
|---|---|---|---|
| $j_o$ in insert (Bi2212) | A/mm$^2$ | 570 | 570 |
| $j_o$ in outsert (Nb$_3$Sn) | A/mm$^2$ | 570 | 380 |
| Coil width insert/outsert | mm | 47/22 | 33/55 |
| Area quadrant coil insert/outsert | mm$^2$ | 2387/1912 | 1434/4924 |
| $B_{bore}$ stand-alone insert/outsert | T | 16.0/10.7 | 12.5/14.1 |
| $\sigma_\vartheta$ hybrid | MPa | -191 | -177 |
| $\sigma_r$ hybrid | MPa | -212 | -196 |
| $\sigma_\vartheta$ stand-alone insert/outsert | MPa | -144/-249 | -108/-175 |
| $\sigma_r$ stand-alone insert/outsert | MPa | -142/-85 | -93/-126 |
| $E$ hybrid (4 quad) | MJ/m | 2.3 | 2.6 |
| $E$ stand-alone insert/outsert (4 quad) | MJ/m | 1.0/1.4 | 0.5/2.47 |

In order to reduce the amount of HTS material, in Case II we imposed a lower $j_o$ to the Nb$_3$Sn with respect to the Bi2212 (380 vs 570 A/mm$^2$) coils and we re-optimized the dimension to re-obtain 20 T bore field with a 15% margin in both HTS and LTS coils. The result is a significant increase in Nb$_3$Sn coil, but with a substantial reduction in Bi2212 coil. This is an interesting result if the goal is to minimize the amount of HTS. This means that if with an increase of the $j_o$ in the outer part of the coil (*grading*) one obtains an overall coil size reduction, in a hybrid situation one can follow an opposite approach (a sort of *anti-grading*) to minimize the amount of HTS material.

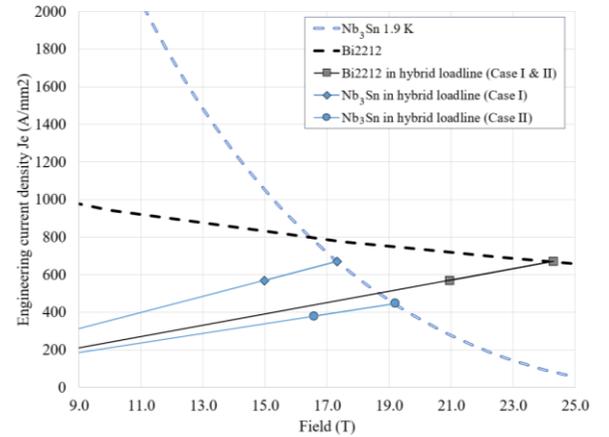

Fig. 5. Critical engineering current density ($j_e = I_{strand}/A_{strand}$) for Nb$_3$Sn and Bi2212, and sector coil load-lines (engineering current density vs. coil peak field) for the two cases analyzed in Fig. 4.

If now the accumulated stress due to the azimuthal and radial electro-magnetic (e.m.) forces are considered we can see in Table III that in in both hybrid magnets the stresses exceed the 180 MPa limit. Interestingly, with such wide coil, the radial stress appears to be the most critical, with generally smaller values for

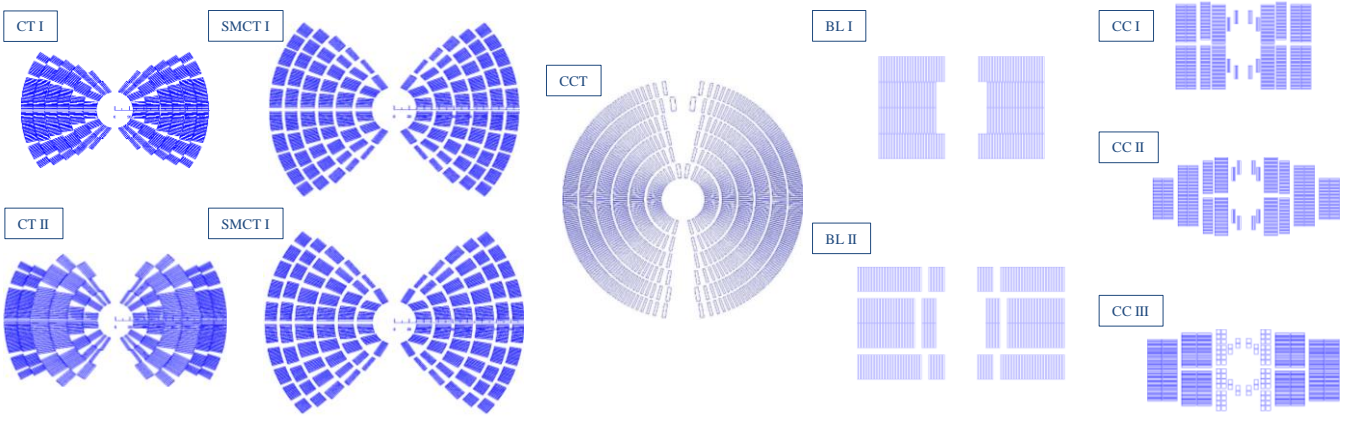

Fig. 6. Cross-sections of 20 T hybrid dipole coils. From left two right: Cos-theta (CT) design, with 4 (top) and 2 (bottom) layer Bi2212 coils; Stress management Cos-theta (SMCT) design, with 4 (top) and 2 (bottom) layers Bi2212 coils; Canted Cos-theta (CCT) design, with 4-layer Bi2212 coil; Block (BL) design, with and without stress management; Common Coil (CC) design, with Bi2212 (top, with 3 external $Nb_3Sn$ layers, and center, with 5 external $Nb_3Sn$ layers) and REBCO CORC coils (bottom, with 4 external $Nb_3Sn$ layers). For all the CC designs, only one aperture is shown.

TABLE IV
20 T HYBRID MAGNET PARAMETERS

| Parameter | Unit | CT I | CT II | SMCT I | SMCT II | CCT | BL I | BL II | CC I | CC II | CC III |
|---|---|---|---|---|---|---|---|---|---|---|---|
| Ins. cable I width/thick. | mm | 10.7/1.5 | 12.0 / 1.7 | 12.3 / 1.5 | 12.3 / 1.5 | 18.7/1.9 | 17.1/2.1 | 17.1/2.1 | 18.7 / 1.8 | 18.7 / 1.8 | 7.5 / 7.5 |
| Ins. cable II width/thick. | mm | 9.4/1.5 | 14.2 / 2.1 | 10.7 / 1.5 | 13.9 / 1.5 | - | 17.1/2.1 | 17.1/2.1 | 13.6 / 1.9 | 13.6 / 1.9 | 21.6 / 1.9 |
| Ins. cable III width/thick. | mm | 9.3/1.5 | 7.9 / 1.5 | 9.1 / 1.5 | 9.1 / 1.5 | - | - | - | - | - | - |
| Current_op | kA | 10.7 | 13.0 | 11.4 | 11.8 | 12.8 | 12.6 | 12.2 | 14 | 13.9 | 17.8 |
| B_bore_op | T | 20.0 | 20.0 | 20.1 | 20.0 | 20.0 | 20.0 | 20.0 | 20.0 | 20.0 | 20.0 |
| B_peak_op HTS/LTS | T | 20.5 / 12.7 | 20.3 / 16.1 | 20.6 / 13.6 | 20.6 / 16.0 | 20.2 / 13.2 | 20.6 / 15.1 | 20.9 / 15.2 | 20.4 / 13.8 | 20.2 / 13.7 | 21.0 / 17.0 |
| B_bore_ss | T | 24.4 | 23.5 | 24.4 | 23.2 | 23.4 | 23.6 | 23.6 | 22.9 | 23 | 21.7 |
| B_peak_ss HTS/LTS | T | 24.9/15.4 | 23.8 / 17.7 | 24.9/ 16.4 | 23.8 / 18.4 | 23.6 / 12.9 | 24.3 / 17.7 | 24.7 / 18.0 | 23.3 / 15.7 | 23.3 / 15.7 | 24.7 / 18.2 |
| Load-line margin | % | 18 / 25 | 21 / 15 | 22 / 18 | 20 / 15 | 14 / 14 | 21/17 | 22 / 17 | 13 / 13 | 13 / 13 | 15 / 7 |
| Area quad. ins. cable HTS | mm$^2$ | 3241 | 1494 | 2091 | 1527 | 4490 | 1360 | 1500 | 1290 | 1154 | 1012 |
| Area quad. ins. cable LTS | mm$^2$ | 2150 | 6106 | 3780 | 5148 | 4915 | 4740 | 4640 | 2326 | 2558 | 4191 |
| Coil width* | mm | 105 | 129 | 144 | 149 | 135 | 80 | 112 | 70 | 104 | 106 |
| Coil inner radius* | mm | 25 | 25 | 30 | 30 | 30 | 35 | 35 | 25 | 25 | 25 |

\* (Inner radius innermost cable on the mid-plane) - (Outer radius of outermost cable on the mid-plane)
\*\* Inner radius innermost cable on the mid-plane

Case II. Also, when tested in stand-alone, one can notice the extremely high stress in the large aperture $Nb_3Sn$ coils.

Finally, the stored energy for the 20 T hybrids and the stand-alone cases are provided in Table III: also for these parameters, the most critical condition also appears to be the one of the large aperture $Nb_3Sn$ coil, with a stored energy similar to the 20 T hybrid case.

## V. MAGNETIC DESIGN

For a preliminary magnetic analysis of the 20 T hybrid magnet, we investigated different coil design options, all shown in Fig. 6. In particular, the following lay-outs were considered: traditional Cos-theta (CT) design, Stress Management Cos-ϑ (SMCT) design, Canted Cos-theta (CCT) design, Block (BL) design, and Common-Coil (CC) design. The main parameters are given in Table IV. The cables considered range from 9.1 to 21.6 mm width and from 1.5 to 2.1 mm thickness (including insulation). All the designs implement HTS Bi2212 Rutherford cables; REBCO CORC wire, with an assumed 7.5 x 7.5 mm$^2$ of dimension with insulation, was considered only for the common-coil, given the large bending radius of the latter. With a 20 T bore field, the operational current varies from 10.7 to 17.8 kA, and the peak field in the HTS and LTS coils is respectively 20.2-21.0 T and 12.7-17.0 T. The target load-line margin is achieved in the CT, SMCT, and BL designs, while in the others the margin is a few percentage points below target. In terms of field quality, design criteria are met by all the designs except the BL, which features geometric harmonics up to the 10 units level.

Finally, it is important to point out that only a preliminary investigation of the accumulated electro-magnetic (e.m.) forces in some of the designs was carried out, and a complete mechanical analysis aimed at bringing the stress in the HTS and LTS below the limits fixed in Table II has not been performed yet. Therefore, the designs depicted in Fig. 6 represent only a first iteration and a starting point of the design, and, since they meet only part of the criteria, they are not yet comparable. In the next sub-sections, we formulate some initial considerations for each design.

### A. Cos-theta (CT), Stress Management Cos-theta (SMCT), and Canted Cos-theta (CCT) Designs

For the CT options we considered a design with double layers coils, each wound with the same cable. This design choice avoids interlayer splices and has been implemented in most of the Nb$_3$Sn CT coils fabricated so far (the only exception being the CERN-ELIN and UT-CERN dipole magnets [9]). The first design (top CT in Fig. 6) has 4 Bi2212 layers and 2 Nb$_3$Sn layers, with a peak field in the Nb$_3$Sn of 12.7 T and a total coil width of 105 mm. As we did for the sector coils, in a second design (bottom CT in Fig. 6) we reduced the Bi2212 layers from 4 to 2 by increasing the width of the Nb$_3$Sn coils. As a result, the peak field in the Nb$_3$Sn rose to 16.1 T and the total coil width rose to 129 mm. For a detailed description of these options we refer to [36], whereas the description of 4-layer CT option for high-field Nb$_3$Sn magnet can be found in [37]. As shown by the sector coil analysis, a traditional CT design magnet aiming at the 20 T level is characterized by high coil stress both in the azimuthal and in the radial direction. A possible alternative solution is represented by the SMCT, where each layer is separated by 5 mm thick spars (or mandrels) and each cable block is separated by ribs, connected to the mandrel [38]-[40]. The implementation of stress intercepting elements results in an overall increase of coil width from 102-129 mm in the CT to 144-149 mm, and in the conductor area. A further step towards the reduction of the stress is done with the CCT design, where each turn is separated by spars and ribs [41]-[43]. The field quality is naturally achieved by superimposing the two tilted solenoids (see Fig. 6 center). For the 20 T hybrid we chose a simple design with 4 Bi2212 layers and 2 Nb$_3$Sn layers, all wound with a MQXF cable. The total area of the insulated cable (taken from a simple cross-section of the 3D design) is, as expected, larger than in the previous CT and SMCT designs. However, since the layer-to-layer splices are located in the coil ends, a full grading coil, with cables progressively smaller from the inserts to the outsert, can reduce significantly the coil size, and it will be the goal of the next step in the optimization.

### B. Block (BL) Design

The Block design [44]-[46] allows for a very efficient subdivision between the HTS and LTS coils, since the cables are aligned with the flux lines. Therefore, the area of Bi2212 in the block design shown in Fig 6 (BL top design), is smaller than for the CT, SMCT and CCT options (see Table IV). Also, in terms of total conductor area the design is very compacted, despite the inclusion of a 10 mm thick internal support in the inner coil that increases the coil aperture to 70 mm (similarly to the coil design of FRESCA2 [47] and TDF [48] magnets). However, as shown in [49], the peak stress in the coil, in particular because of the horizontal e.m. forces, can be as high as 280 MPa in the Nb$_3$Sn and 160 MPa in the Bi2212 at 20 T. Therefore, an alternative has been considered where vertical and horizontal plates are included to intercept part of the e.m. forces. In the bottom BL design in Fig. 6, intercepting plates separates Bi2212 and Nb$_3$Sn coils: consequently, the coil increases in size, but the stress in the Nb$_3$Sn and in the Bi2212 coils decreases to about 160 MPa and 140 MPa respectively at 20 T. For a complete description of the two designs, and a discussion about fabrication issues and stress management options, we refer to [49]. Further design work on the BL design will be aimed at reducing the coil stress and improve the field quality.

### C. Common-coil design

On the right side of Fig. 6, 3 different common coil designs for the 20 T hybrid are shown. The common-coil design [50]-[52] is based on racetrack coils that, with a very large bending radius in the ends, cover both magnet apertures (in Fig. 6 only one aperture is shown). The large bending radius opens the possibility not only of implementing the *react-and-wind* technique, but also to utilize REBCO CORC cable, whose rigidity makes small bending radius a possible source of conductor degradation. Similar to the block design described above, the common-coil allows aligning the block with the flux lines, thus minimizing the HTS conductor use. Also, by having the layer-to-layer splice inside the winding pole at the center of the coil, one can wind and react individual layers, and "grade" each layer to maximize efficiency. The two top designs in Fig. 6 uses Bi2212 cables in the small blocks around the aperture and in the first layer, followed by either 3 or 5 layers of Nb$_3$Sn cables. In both designs, the coil area and width are small compared to the previous designs; however, it is important to point out that the load-line margins are below the 15% criteria. In the third design, we implement a large CORC wire, in series with 4 layers of HTS. Also in this case further magnetic analysis will be carried out to bring the load-line margin to the design criteria. Regarding the coil stress, the common coil design allows the insertion of vertical plates to intercept the horizontal e.m. forces, and similarly to the BL design, horizontal bars can be considered to intercept the vertical force. As a next step, a mechanical analysis will be performed to verify the stress, and the magnetic design will be updated accordingly.

## VI. CONCLUSIONS

We presented in this paper a preliminary investigation of a hybrid 20 T dipole, which we consider a promising option for a dipole magnet operating beyond the limits of Nb$_3$Sn and aimed at minimizing the HTS volume. Two HTS conductors are considered: Bi2212, in the form of a Rutherford cable with $j_e$ of 740 A/mm$^2$ at 20 T, and REBCO tape in a CORC/STAR wire with $j_e$ of 590 A/mm$^2$ at 20 T. As part of the design criteria, we target a bore field of 20 T with a load-line margin of at least 15% for both LTS and HTS coils. Also, all the coils shall be powered in series, and stress must be kept below 150-180 MPa in the Nb$_3$Sn and below 120 MPa in the HTS. A preliminary analysis done with sector coils indicated that 1) with identical $j_o$ in both HTS and LTS, we have a coil width of ~70 mm, 2) radial stresses of about 200 MPa are generated by the radial/horizontal e.m. forces, and 3) a significant reduction of HTS area can be obtained by "anti-grading", i.e. by increasing the size of the Nb$_3$Sn outsert. Finally, we performed a preliminary analysis of a 20 T hybrid with different coil design options, all shown in Fig. 6. The designs are not comparable yet since they do not meet all the specifications, but they provide a first idea of the overall coil features, and they constitute a starting point for further analysis. The next step will include a mechanical analysis, and the continuation of the magnetic analysis, with the goal of meeting margin, field quality, and stress criteria in all the designs.